# Strong Exciton-Phonon Coupling as a Fingerprint of Magnetic Ordering in van der Waals Layered CrSBr

Kaiman Lin[1,2], Xiaoxiao Sun[2], Florian Dirnberger[3], Yi Li[2,4], Jiang Qu[5], Peiting Wen[2,4], Zdenek Sofer[6], Aljoscha Söll[6], Stephan Winnerl[2], Manfred Helm[2,4], Shengqiang Zhou[2], Yaping Dan[1,*], Slawomir Prucnal[2,*]

[1] University of Michigan-Shanghai Jiao Tong University Joint Institute, Shanghai Jiao Tong University, 20024 Shanghai, P. R. China

[2] Helmholtz-Zentrum Dresden-Rossendorf, Institute of Ion Beam Physics and Materials Research, Bautzner Landstrasse 400, 01328 Dresden, Germany

[3] Institute of Applied Physics and Würzburg-Dresden Cluster of Excellence ct.qmat, Technische Universität Dresden, Germany

[4] Technische Universität Dresden, 01062 Dresden, Germany

[5] Leibniz Institute for Solid State and Materials Research Dresden (IFW Dresden), Helmholtzstraße 20, 01069 Dresden, Germany

[6] Department of Inorganic Chemistry, University of Chemistry and Technology Prague, Technická 5, 16628 Prague 6, Czech Republic

[*]corresponding authors: s.prucnal@hzdr.de, yaping.dan@sjtu.edu.cn

**ABSTRACT:** The layered, air-stable van der Waals antiferromagnetic compound CrSBr exhibits pronounced coupling between its optical, electronic, and magnetic properties. As an example, exciton dynamics can be significantly influenced by lattice vibrations through exciton-phonon coupling. Using low-temperature photoluminescence spectroscopy, we demonstrate the effective coupling between excitons and phonons in nanometer-thick CrSBr. By careful analysis, we identify that the satellite peaks predominantly arise from the interaction between the exciton and an optical phonon with a frequency of 118 cm$^{-1}$ (~14.6 meV) due to the out-of-plane vibration of Br atoms. Power-dependent and temperature-dependent photoluminescence measurements support exciton-phonon coupling and indicate a coupling between magnetic and optical properties, suggesting the possibility of carrier localization in the material. The presence of strong coupling between the exciton and the lattice may have important implications for the design of light-matter interactions in magnetic semiconductors and provide insights into the exciton dynamics in CrSBr. This highlights the potential for exploiting exciton-phonon coupling to control the optical properties of layered antiferromagnetic materials.

Graphic abstract

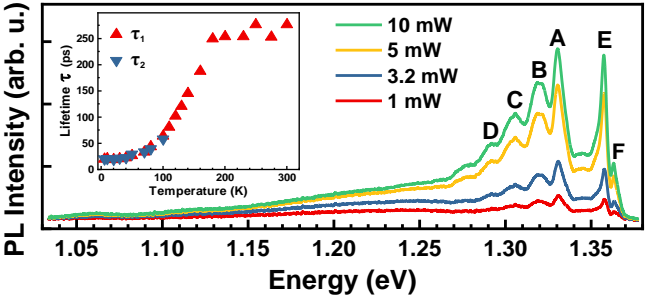

Exciton-phonon coupling,[1] the interaction between excitons and lattice vibrations (phonons), is pivotal in determining the optical and electronic properties of materials, providing valuable insights into light-matter interactions. Various techniques, such as photoluminescence (PL) spectroscopy, absorption spectroscopy, and reflection spectroscopy, offer avenues for investigating the intricate interplay between excitons and phonons.[2–7] In conventional transition metal dichalcogenides, such as 2D $MoS_2$, the existence of exciton-phonon coupling has been experimentally demonstrated by resonant Raman scattering and theoretically verified.[8,9] CrSBr is an A-type antiferromagnetic (AFM) material consisting of van der Waals (vdWs) ferromagnetic monolayers that are antiferromagnetically coupled along the stacking direction.[10,11] Unlike most magnetic materials, CrSBr is a direct band gap semiconductor, providing opportunities for exploring the interplay between magnetic, optical, and electrical properties. It possesses a sizable band gap of approximately 1.5 eV[10,12] and demonstrates good air stability.[10,13] Moreover, it exhibits a high transition temperature for antiferromagnetic coupling, reaching 132 K for bulk CrSBr.[11,14,15] These characteristics position CrSBr as a highly promising material for applications in optoelectronics, spintronics, and quantum technologies.[16–19]

Recent studies[20–23] on the band structure of CrSBr have revealed that the conduction band along the $\Gamma-X$ direction is nearly flat, suggesting a high likelihood of phonon-assisted transitions. Using micro-Raman scattering, strong spin-phonon coupling was observed in bulk CrSBr.[23] Moreover, Klein *et al.*[21] have shown that the bulk CrSBr hosts quasi-1D excitons with strong quasiparticle interactions. Torres *et al.*[13] have shown that spin-phonon coupling in a few-layers-thick CrSBr can be modified by He-ion irradiation. Interestingly, thin film CrSBr exhibits ferromagnetic phases at low temperatures, with a Curie temperature of approximately 40 K.[10] The underlying cause of this ferromagnetism remains unclear. However, our recent studies[24] suggest that it is mediated by defects, which can be controllably generated by ion irradiation.

Wilson *et al.*[20] recently observed low-temperature PL peaks at around 1.34 eV and 1.36 eV in few-layer CrSBr. They attributed the lower energy peak to the direct radiative recombination of excitons near the bandgap energy, and the high energy peak was ascribed to the transition between the second conduction band minimum and the valence band maximum at the $\Gamma$-point. Interestingly, in thicker flakes, the PL emission showed exciton-related peaks within the range of 1.33 to 1.37 eV, with variations depending on the measurement geometry and film thickness.[25] In addition to the pronounced excitonic signatures of few-layer flakes,[26–28] a recent study[29] found strong exciton-photon coupling to play a distinctive role in the optical properties of nanometer-thick CrSBr.

In this work, we identify strong exciton-phonon coupling in nanometer-thick CrSBr using temperature-dependent PL spectroscopy and exciton lifetime measurements. The presence of exciton-phonon coupling is directly observed in the PL spectrum at 4 K, which displays a distinctive periodic pattern.

Through an in-depth analysis of this periodic pattern, as well as an examination of its power and temperature dependency, we gain valuable insights into the intricate interplay between exciton and phonon coupling as the temperature is varied. Moreover, by investigating the integrated PL intensity, peak energy, and lifetime as functions of temperature, we propose that the radiative recombination in CrSBr might be governed by carriers localized at discrete energy states, resembling those observed in quantum well systems, due to temperature-dependent intralayer ferromagnetic coupling between Cr atoms and interlayer antiferromagnetism.[30–32]

**RESULTS AND DISCUSSION**

Mechanically exfoliated CrSBr flakes were transferred onto a 260 nm $SiO_2$/Si substrate (see Methods). Figure 1a exhibits an optical microscope image of a typical CrSBr flake investigated in this paper. In this particular case, the thickness measured by atomic force microscopy (AFM) is 102 nm (Figure 1b). Figure 1c displays the Raman spectra obtained with an excitation laser energy of $E_L$=2.33 eV (λ=532 nm) and polarization along a and b axis, respectively. Structurally, CrSBr crystallizes similarly to FeOCl layered magnets, adopting an orthorhombic crystal structure within a *Pmmn* space group and a $D_{2h}$ point group.[13,33–36] Within the first Brillouin zone, CrSBr is predicted to exhibit 18 phonon modes, including 15 optical modes and 3 acoustical branches.[23] In back scattering geometry, with incident laser light perpendicular to the sample surface, the room temperature Raman spectra of CrSBr reveal only three main optical phonons of $A_g$ symmetry. Here, we observe out-of-plane $A_g^1$ mode ~ 114 cm$^{-1}$, $A_g^2$ mode ~ 244 cm$^{-1}$ and $A_g^3$ mode ~ 344 cm$^{-1}$ with laser polarization along b axis, and $A_g^2$ and $A_g^3$ modes with laser polarization along a axis, due to the anisotropic structural morphology of the crystal.[13] Other phonon modes become visible at room temperature under varying angle excitation; for example, three $B_{2g}$ modes at the Γ-point were detected upon tilting the sample by 60º with respect to the laser light.[23] Regarding PL measurements, non-resonant micro-PL excitation was employed, again utilizing a laser energy of 2.33 eV. The room temperature PL spectrum presented in Figure 1d shows a single peak at ~1.28 eV, which can be attributed to radiative recombination of excitons in CrSBr.[28]

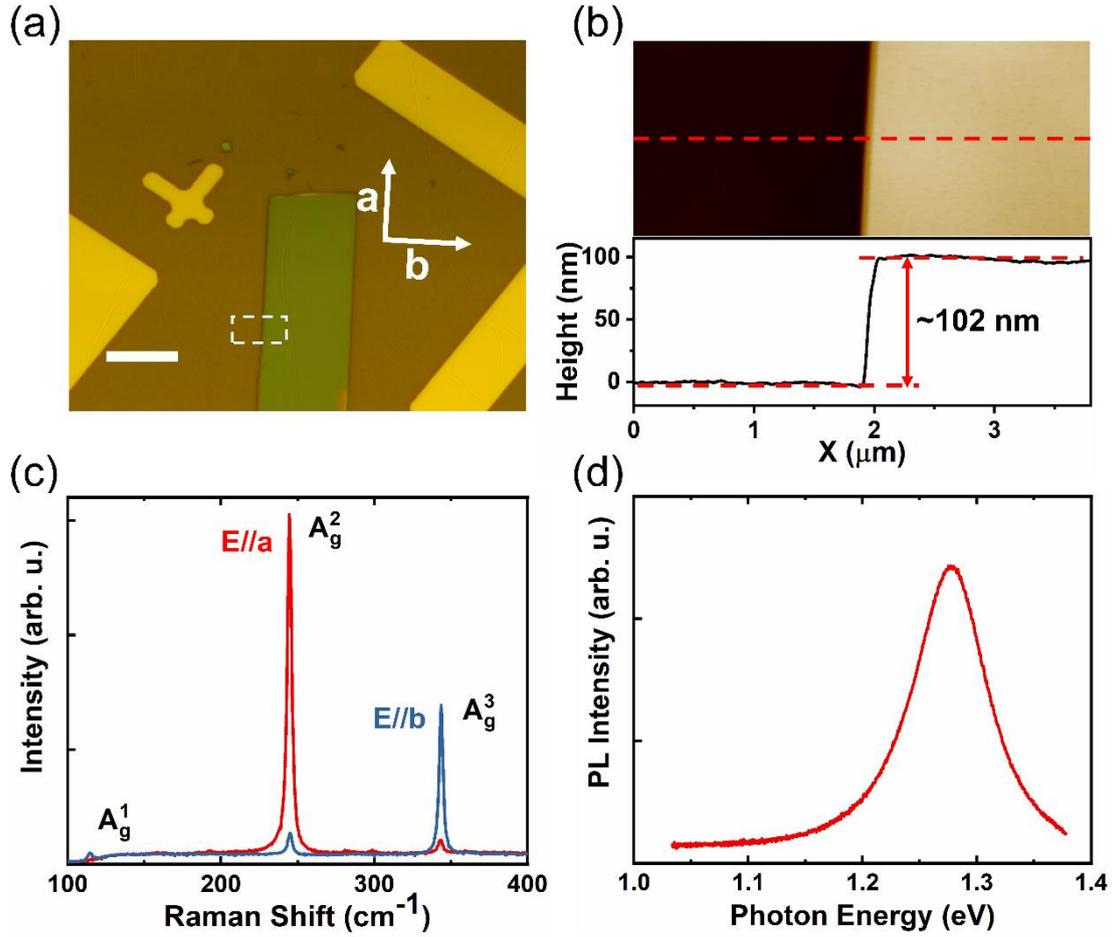

**Figure 1. Properties of thin CrSBr flake.** (a) Optical microscopic image of a CrSBr flake on a Si substrate with 260 nm thick $SiO_2$. Scale bar, 5 μm. (b) Atomic force microscopic image and corresponding height profile of the area enclosed by the dashed rectangle in (a). (c) Raman and (d) PL spectra under 2.33 eV laser excitation at room temperature. The Raman spectra are measured with excitation laser polarized along a and b axis, respectively, showing the out-of-plane $A_g^1$ mode ~ 114 $cm^{-1}$, $A_g^2$ mode ~ 244 $cm^{-1}$ and $A_g^3$ modes ~ 344 $cm^{-1}$.

Figure 2a presents a representative PL spectrum acquired at 4 K for the 102 nm thick CrSBr flake. The spectrum displays its highest emission intensity at approximately 1.33 eV, marked as peak A. Intriguingly, we observed satellite peaks resembling phonon sidebands on the low energy side of peak A in the PL spectrum. To accurately determine the distance between the emission peaks, the PL spectrum was fitted with Lorentzian functions. The fitting analysis revealed that the individual peaks were separated by approximately the same energy of 14.3±1.3 meV (115±9 $cm^{-1}$), indicating the involvement of phonons in the PL process. Particularly, this energy value is in reasonable agreement with the phonon energy of the $A_g^1$ phonon mode (14.6 meV) obtained from the Raman spectrum measured at 4 K (Figure S2). Therefore, we assigned the low energy peaks to the phonon sidebands arising from exciton-phonon coupling within our CrSBr system. Figure 2b displays the power-dependent PL spectra obtained at low temperature (4 K). The most prominent peaks in the spectrum are labeled as peak A to F. Notably, as the pumping power increases, the normalized intensities and positions of all peaks remain unchanged (Figure S3). This persistent behavior indicates that peaks A to F share a common origin, which is likely related to the excitonic properties of the CrSBr system.

To further understand the relationship between identified peaks as well as the exciton recombination process in the CrSBr material, we conducted the same Lorentzian fitting procedure on the power-dependent PL spectra as shown in Figure 2a. Figure 2c presents double-logarithmic plots of the PL intensity ($I$) of peaks A to D versus excitation power ($L$). The curves can be fitted by a simple power law

$$I \propto L^k, \tag{1}$$

where $L$ represents the excitation power and $k$ is a dimensionless exponential coefficient. Importantly, we observed no saturation effects within the investigated power range and a nearly linear dependence of PL intensity on the excitation power. The exponential coefficient $k$ can be utilized to determine the deexcitation process in the material. When $k$ exceeds 1, it suggests stimulated emission, whereas a value smaller than one indicates the presence of non-radiative channels, such as defects, Auger recombination, or exciton-to-trion conversion, commonly observed in 2D vdWs crystals.[37–40] For our investigated peaks A, B, C, and D, the obtained exponent $k$ is approximately 1±0.1, which is typical for radiative recombination of neutral excitons. The similarity in these exponent values further reinforces the understanding that peaks B to D are phonon sidebands associated with exciton-phonon coupling states in the CrSBr material.

Similarly, Figure 2d shows the PL intensity ($I$) of peaks A, E, and F plotted as a function of the excitation power ($L$) using the same power law fitting. The exponent value $k$ obtained for this set of peaks is also close to 1±0.1, suggesting that peaks E and F are also due to the radiative recombination of excitons. Recently, Dirnberger et al.[29] showed that CrSBr crystals may support strong exciton-photon coupling even in the absence of external, highly reflective cavity mirrors. The high energy emission is in good agreement with the exciton-polariton state expected to emerge from the strong coupling of excitons and photons confined in our 102 nm-thick CrSBr crystal. As expected for this coupling, the peak position varies as the thickness of CrSBr changes (Figure S1). Additionally, some contributions may also result from transitions between the second conduction band minimum and the valence band maximum at the Γ-point, as proposed by Wilson et al.[20].

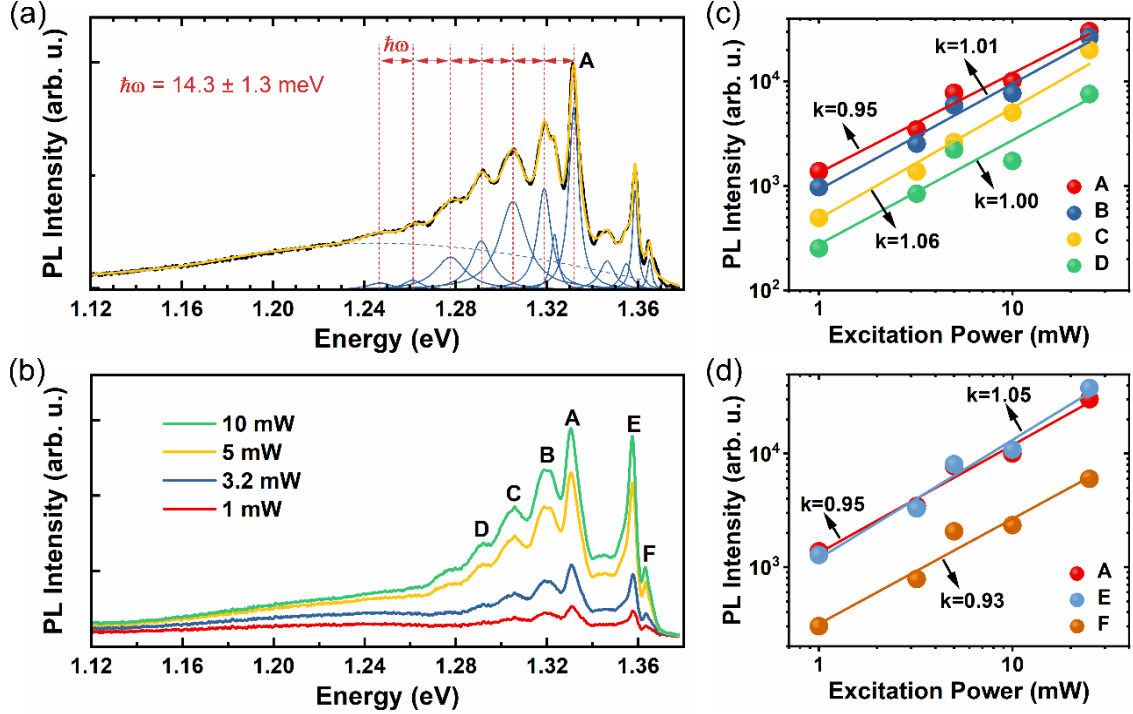

**Figure 2. μ-PL spectrum of thin CrSBr flake obtained at 4 K.** (a) PL spectrum acquired at 4 K with excitation power at 3.2 mW of a 102 nm thick CrSBr crystal on a Si substrate with 260 nm thick $SiO_2$. Black line: PL spectrum, Yellow line: sum of individual peaks, Blue lines: individual peaks. (b) Power-dependent PL spectra at 4 K. (c) PL intensity of the A, B, C and D peaks versus excitation power at 4 K. (d) PL intensity of the A, E and F peaks versus excitation power at 4 K. The experimental points in (c) and (d) were obtained by decomposing the photoluminescence spectra in a series of Lorentzians.

To further corroborate the excitonic origin of the presented PL spectra, we have analyzed the temperature-dependent peak positions for emissions A, E, and F (Figure S4a). Evidently, the change in peak position as a function of temperature shows a consistent behavior for all three peaks, providing strong support for the hypothesis that they indeed have the same origin. Figure S4b displays the energies of peaks A to F across different thicknesses of CrSBr. For peaks A to D, the peak positions remain relatively unchanged regardless of the thicknesses, indicating the presence of exciton-phonon coupling in CrSBr. Conversely, peak E exhibits variations in its peak positions for different thicknesses, as expected for exciton-polariton states.[29]

To explore the dynamics of exciton-phonon coupling with respect to the temperature, we conducted temperature-dependent PL measurements on the same sample. Figure 3a illustrates the evolution of the PL spectrum as the temperature increases from 4 K to 120 K. Notably, within the temperature range of 4 K to 60 K, distinct and prominent phonon replicas are observable. However, as the temperature rises, these phonon replicas gradually lose their individual prominence and merge into a single broad peak. This merging phenomenon arises from the thermal broadening of the phonon energy distribution.[41] Figure 3b provides a comprehensive view of this blending effect by displaying the normalized PL spectra from 4 K to 300 K. Below $T_N$, the temperature-dependence of peaks E and F shown in Figure S4a is mainly determined by incoherent magnons, *i.e.*, thermal fluctuations of the magnetic order.[29] In general, the position of the PL peak in energy and the PL intensity are therefore fairly complex and must include interaction of excitons with photons, phonons and magnons. Above $T_N$, however, exciton-magnon coupling can be neglected and the change of the PL energy and intensity is dominated by exciton-phonon coupling.

In our study on the dynamics of exciton-phonon coupling, we plotted the Lorentzian-fitted peak energy, full width at half maximum (FWHM) and integrated PL intensity as a function of temperature, as depicted in Figure 3c. The details of the fitting procedure are outlined in Supporting Information (SI) and Figure S5. It is noteworthy that our fitting and integration involve only the exciton-phonon coupling branch.

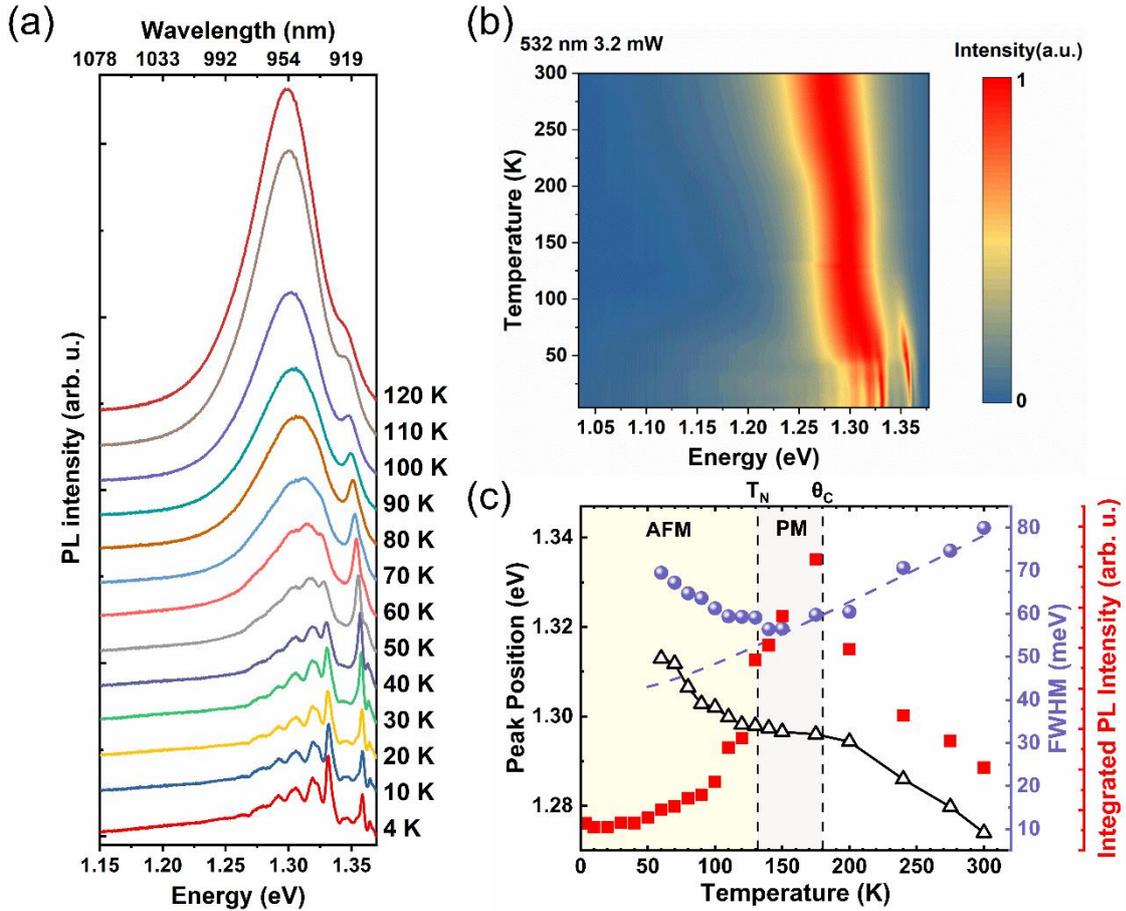

**Figure 3. Temperature-dependent µ-PL of thin CrSBr flake.** (**a**) Evolution of PL spectrum measured from T=4 K to T =120 K. (**b**) Color map of normalized PL spectra. (**c**) Peak position (triangles), FWHM (circles) and integrated PL intensity (squares) of the exciton-phonon coupling branch as a function of temperature. The blue dashed line is the fitting curves used to determine the Fröhlich interaction of excitons with longitudinal-optical phonons (LO). $T_N$ and $\theta_C$ stand for the Néel temperature and paramagnetic Curie temperature, respectively.

Regarding the peak energy, we observed a consistent red shift with increasing temperature from 4 K to 300 K. Notably, the shift of the PL emission peak demonstrates a distinct vertically mirrored and tilted S-shaped pattern. Typically, in semiconductors, the temperature-dependent peak position exhibits a monotonic red-shift with increasing temperature,[42] and this behavior can be well described by the Varshni model, which provides valuable insights into the band structure of semiconductors. In classical semiconductors, the main factor influencing the change in the band gap with temperature is the temperature-dependent electron-lattice interaction. This interaction causes a shift in the relative positions of the conduction and valence bands. Additionally, the temperature-dependent lattice dilatation also contributes to this phenomenon. The observed anomaly of the peak position shift as a function of temperature has been previously reported in quantum wells and certain bulk III-V compound semiconductors, such as GaN, AlGaN, and InGaN.[30–32,43,44] The so-called S-shaped temperature dependence of the PL peak energy is attributed to exciton localization in band-tail states that are formed due to potential fluctuations caused by variation of alloy composition within quantum wells and layer thickness.[45] In the case of layered magnetic materials CrSBr, potential fluctuations can arise from various sources, including the changes in magnetic coupling and the existence of strongly localized states. Our investigation reveals significant deviations in the temperature-dependent behavior within distinct temperature ranges: from about 60 K to about 120 K, from about 120 K to about 180 K, and above 180 K to 300 K, which coincide with different magnetic phase regimes in CrSBr. A similar trend is also observed in the change of the integrated PL intensity as a function of temperature. Specifically, below 40 K, the integrated PL intensity remains relatively stable, but as the temperature increases from 40 K up to approximately 180 K, the integrated PL intensity also increases. Beyond 180 K, the integrated PL intensity starts to decrease. At low temperatures, CrSBr exhibits strong intralayer ferromagnetic coupling between Cr atoms and weak interlayer antiferromagnetic coupling.[10,35,46] Moreover, the negative magnetoresistance measurements reported by Telford *et al.*[10] suggest the existence of ferromagnetic (FM) states below 40 K and a transition from the antiferromagnetic (AFM) to the paramagnetic (PM) phase in the temperature range of 120 K to 150 K, with a Néel temperature of approximately 131 K, and short-range correlations observed up to 185 K.[29] Significantly, the threshold temperatures for the AFM and PM states coincide with the changes in PL emission as a function of temperature in the investigated CrSBr material. Based on these observations, we propose that the S-shaped dependence of the PL peak energy on temperature originates from competing effects of magnons and phonons on exciton energies.[29,33] Therefore, the shift in the PL peak position provides a valuable approach to investigate the magnetic properties of CrSBr at the micro- or even nanoscale, which can be challenging using conventional magnetometry techniques.

Furthermore, the FWHM of the PL spectrum also exhibits unusual behavior as a function of temperature, similar to the integrated PL intensity and peak position. In general, the FWHM of PL emission line as a function of temperature can provide information about the exciton-phonon coupling in semiconductors. In most typical semiconductors, as the temperature increases, the FWHM tends to broaden. However, in our case, we observe a U-shaped trend in the FWHM as a function of temperature. Specifically, the FWHM decreases with increase of temperature up to approximately 130 K and then starts to increase. This distinctive behavior is reminiscent of the U-shaped FWHM observed in GaN-based light-emitting devices, attributed to localized states.[47] In GaN, at low temperatures, carriers tend to redistribute in shallow and deep localized states, leading to a narrowing of the PL emission line. As the temperature increases, the degree of carrier localization diminishes, resulting in a gradual broadening of the emission line. Moreover, at higher temperatures, the broadening of the FWHM can be attributed to the coupling of excitons with longitudinal phonons (LO). In our case, we observed the smallest FWHM at around 140 K, which matches the Néel temperature and the magnetic phase change from AFM to PM. The FWHM as a function of temperature can be fitted with the formula:

$$\Gamma(T) = \Gamma_0 + \frac{\Gamma_{L0}}{e^{\frac{\omega_{L0}}{k_B T}} - 1}, \qquad (2)$$

where $\Gamma_0$ represents the temperature-independence inhomogeneous broadening term, $\Gamma_{L0}$ is the Fröhlich coupling strength, $\omega_{L0}$ is longitudinal optical phonon energy, $T$ is the temperature, $k_B$ is the Boltzmann constant ($8.617 \times 10^{-2}$ meV/K).[48]

The change in the slope of FWHM versus temperature occurs at about 140 K, indicating the presence of additional scattering centers below $T_N$. In the AFM phase regime, scattering of excitons with incoherent magnons may further contribute to the broadening of the FWHM.[29,33] Above 140 K, the estimated values for $\Gamma_0$, $\Gamma_{L0}$ and $\omega_{L0}$ are approximately 42 meV, 26 meV, and 14 meV, respectively. Importantly, the estimated energy of the longitudinal optical phonon $\omega_{L0}$ agrees well with the phonon energy of $A_g^1$ (118 cm$^{-1}$), providing support for our assumptions about exciton-phonon coupling based on the photoluminescence results (at the current stage, theoretical modelling of the observed phenomenon is beyond our competence). In the multilayer structure of CrSBr, the Br atoms are located at the top and bottom of each individual layer, leading to vibrational interlayer coupling mediated by the Br atoms.[13] Additionally, the Br-related phonon mode exhibits stronger interlayer characteristics compared to the other two phonon modes, $A_g^2$ (Cr-244 cm$^{-1}$) and $A_g^3$ (S-344 cm$^{-1}$), making it more sensitive to changes in the magnetic phases of CrSBr, especially the antiferromagnetic (AFM) coupling between layers.[13]

To gain deeper insights into the underlying mechanism, we employed temperature-dependent time-resolved spectroscopy. Figure 4a and 4b present spectrally-resolved streak-camera images at temperatures of 5 K and 200 K, respectively. At 5 K, the presence of two distinct spectral branches confirms the coexistence of exciton-phonon coupling and exciton-photon coupling. However, at 200 K, only a single branch is observable. In addition, Figure 4c displays representative normalized temperature-dependent decay curves for the spectrally integrated luminescence corresponding to the exciton-phonon coupling branch. From these curves, PL lifetimes for exciton-phonon coupling emission, denoted as $\tau_1$, can be extracted. The same procedures were conducted for the exciton-photon coupling branch, from which the PL lifetimes of exciton-photon coupling emission, $\tau_2$, can be obtained. The evolution of $\tau_1$ and $\tau_2$ as the temperature increases from 5 K to 300 K is depicted in Figure 4d. A notable observation is that below 40 K, $\tau_1$ and $\tau_2$ exhibit a perfect overlap and remain stable. As the temperature increases up to 110 K, the lifetimes for both components increase, yet they remain almost the same, suggesting similar lifetimes for exciton-phonon and exciton-polariton states. However, above 110 K, we could not resolve the $\tau_2$ component. Between 100 K and around 180 K, there is a nearly linear increase in $\tau_1$, followed by a plateau from 180 K to 300 K. The increase in lifetime from 4 K to 180 K and then plateauing also indicates the possibility of carrier localization in thin CrSBr due to the magnetic phases.[49,50] Streak camera images for seventeen different temperatures between 10 K and 300 K are presented in Figure S6.

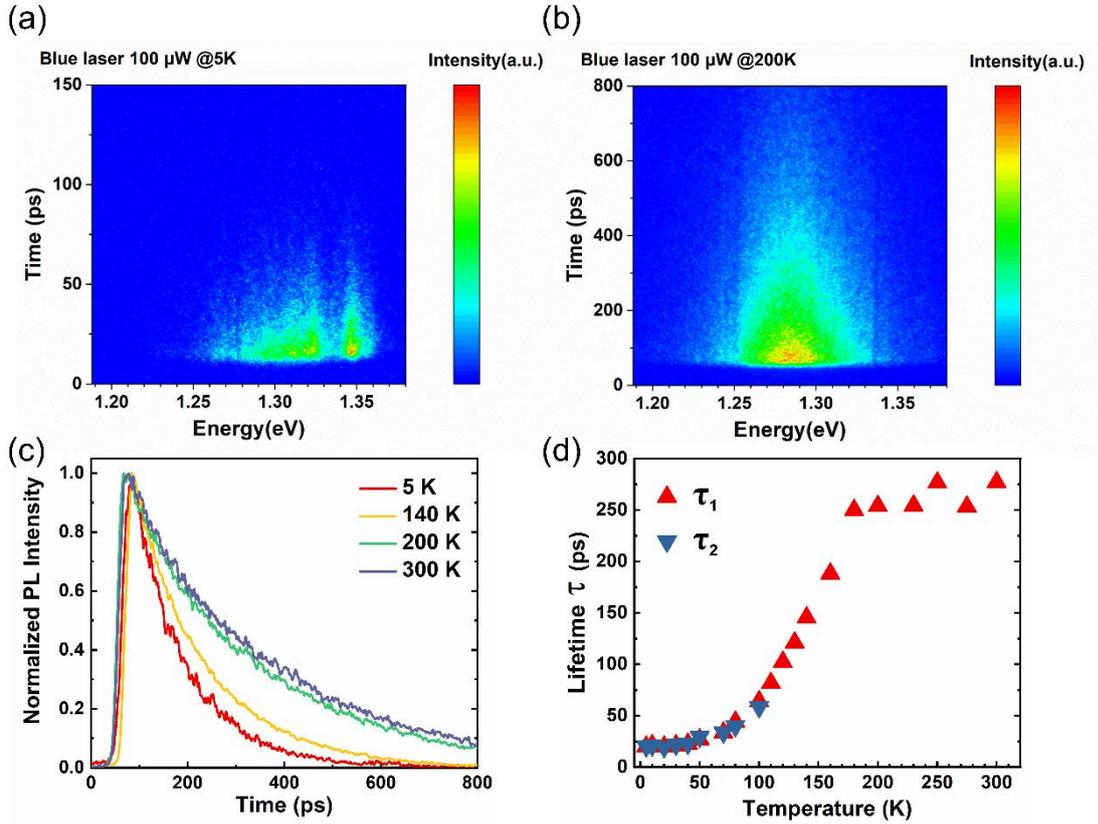

**Figure 4. Temperature-dependent transient μ-PL of thin CrSBr.** (**a,b**) Spectrally-resolved streak-camera image of the PL at temperatures of 5 K and 200 K, respectively. (**c**) Representative normalized temperature-dependent decay curves for the spectrally integrated luminescence corresponding to the exciton-phonon coupling branch. (**d**) Temperature-dependent PL lifetimes of exciton-phonon coupling emission and exciton-photon coupling emission measured under excitation at 400 nm.

**CONCLUSIONS**

To conclude, our photoluminescence studies on CrSBr confirm a strong coupling between lattice vibrations and excitons, mediated by magnetism. The observed S-shaped dependence of the PL peak energy with temperature allows for effective monitoring of magnetic properties in antiferromagnetic layered vdWs crystals, attributable to possible localized carriers arising from magnetic phase transitions. The ability to monitor magnetic phases using light in this kind of materials holds promise for the development of devices and technologies in the fields of spintronics and magneto-optics.

**METHODS**

**Crystal Growth and Sample Fabrication.** CrSBr bulk single crystal was synthesized by a chemical vapor transport method. Chromium (99.99%, -60 mesh), sulfur (99.9999%, 1-6mm), and bromine (99.9999%) elements with a stoichiometry of 1:1:1 were combined and sealed in a quartz ampoule under high vacuum keeping charge in liquid nitrogen batch to avoid bromine evaporation. The material was first pre-reacted in an ampoule at 700°C for 20 hours keeping second end of ampoule under 200°C. The reacted mixture was placed in two-zone horizontal furnace. First the source and growth ends were kept at 850 and 900°C, respectively. After 25 h, the temperature gradient was reversed, and the hot end gradually increased from 880 to 930°C for 10 days period. The high-quality CrSBr single crystals were removed from the ampule in an Ar glovebox. Samples were prepared by mechanical exfoliation using scotch tape method. The substrate used in this paper is silicon with a 260 nm thermal oxide layer. The thicknesses of the exfoliated flakes were confirmed by atomic force microscopy.

**Raman Spectroscopy.** Micro-Raman spectroscopy measurements were performed using a 532 nm excitation laser. The incident beam was focused onto the sample using a 50× objective, with a spot size of ~5 μm in diameter and excitation power of 1 mW. The scattered light was collected by the objective and then dispersed by a Horiba LabRAM HR Evolution Raman spectrometer, and finally detected by a liquid-nitrogen-cooled Charge-Coupled Device (CCD) camera.

**Photoluminescence Spectroscopy.** Micro-PL spectra were acquired using a 532 nm excitation laser. The incident beam was focused onto the sample using a 50× objective, with a spot size of ~5 μm in diameter. A closed-cycle helium cryostat was integrated with the micro-PL system for the temperature-dependent measurements. All thermal cycles were performed at a base pressure lower than $1 \times 10^{-5}$ mbar. The PL signal was detected by a liquid-nitrogen-cooled InGaAs camera.

**Transient Photoluminescence Spectroscopy.** Time-resolved photoluminescence spectra were recorded with a streak camera. The second harmonic of a mode-locked Ti:sapphire (TiSa) laser system at 400 nm with a pulse length of 3 ps was used as excitation source. The spot diameter was around 5 μm. The photoluminescence was collected with a 50× objective and dispersed into a spectrometer (100 grooves mm$^{-1}$), and then coupled into a streak tube and a CCD camera.

## AUTHOR INFORMATION

**Author Contributions**

K.L. prepared the samples, conducted the AFM, Raman, and PL measurements, performed the data analysis and wrote the manuscript; Z.S. and A.S. synthesized CrSBr crystals; X.S. performed the Transient PL measurements; S.W. supported the Raman and PL measurements; Y.L., J.Q. and P.W assisted with measurements. S.Z., M.H. and F.D. discussed the theoretical issues and commented on the manuscript. S.P. and Y.D. supervised and directed the research.

**Notes**

The authors declare that they have no competing interests.

## ACKNOWLEDGMENTS

K. Lin gratefully acknowledges the financial support by Chinese Scholarship Council (File No. 202106230241). Y. Dan acknowledges the financial support by the special-key project of Innovation Program of Shanghai Municipal Education Commission (No. 2019-07-00-02-E00075), National Science Foundation of China (NSFC, No. 92065103), Oceanic Interdisciplinary Program of Shanghai

Jiao Tong University (SL2022ZD107), Shanghai Jiao Tong University Scientific and Technological Innovation Funds (2020QY05). F. Dirnberger gratefully acknowledges financial support from Alexey Chernikov and the Würzburg-Dresden Cluster of Excellence on Complexity and Topology in Quantum Matter ct.qmat (EXC 2147, Project-ID 390858490). Z. Sofer was supported by ERC-CZ program (project LL2101) from Ministry of Education Youth and Sports (MEYS) and used large infrastructure from project reg. No. CZ.02.1.01/0.0/0.0/15_003/0000444 financed by the EFRR. J. Qu gratefully acknowledges financial support from Bernd Büchner (ct.qmat Flex Fund Program: ST 0382022 BB).

## ASSOCIATED CONTENT

**Supporting Information**

PL spectrum of CrSBr with different thickness obtained at 4 K, Raman spectrum for 532 nm excitation at 4 K, Normalized Power-dependent PL spectra for 532 nm excitation at 4 K, Evolution of peak A, E and F from 4 K to 120 K, Energy of peaks in PL spectra obtained at 4 K for CrSBr as a function of thickness, PL spectrum and Lorentzian fitting curves acquired at 60 K, 90 K and 300 K, Fitting Procedure, Spectrally-resolved streak-camera image of the PL at temperatures from 10 K to 300 K.

# Supporting Information

# Strong Exciton-Phonon Coupling as a Fingerprint of Magnetic Ordering in van der Waals Layered CrSBr

Kaiman Lin[1,2], Xiaoxiao Sun[2], Florian Dirnberger[3], Yi Li[2,4], Jiang Qu[5], Peiting Wen[2,4], Zdenek Sofer[6], Aljoscha Söll[6], Stephan Winnerl[2], Manfred Helm[2,4], Shengqiang Zhou[2], Yaping Dan[1,*], Slawomir Prucnal[2,*]

[1] University of Michigan-Shanghai Jiao Tong University Joint Institute, Shanghai Jiao Tong University, 20024 Shanghai, P. R. China

[2] Helmholtz-Zentrum Dresden-Rossendorf, Institute of Ion Beam Physics and Materials Research, Bautzner Landstrasse 400, 01328 Dresden, Germany

[3] Institute of Applied Physics and Würzburg-Dresden Cluster of Excellence ct.qmat, Technische Universität Dresden, Germany

[4] Technische Universität Dresden, 01062 Dresden, Germany

[5] Leibniz Institute for Solid State and Materials Research Dresden (IFW Dresden), Helmholtzstraße 20, 01069 Dresden, Germany

[6] Department of Inorganic Chemistry, University of Chemistry and Technology Prague, Technická 5, 16628 Prague 6, Czech Republic

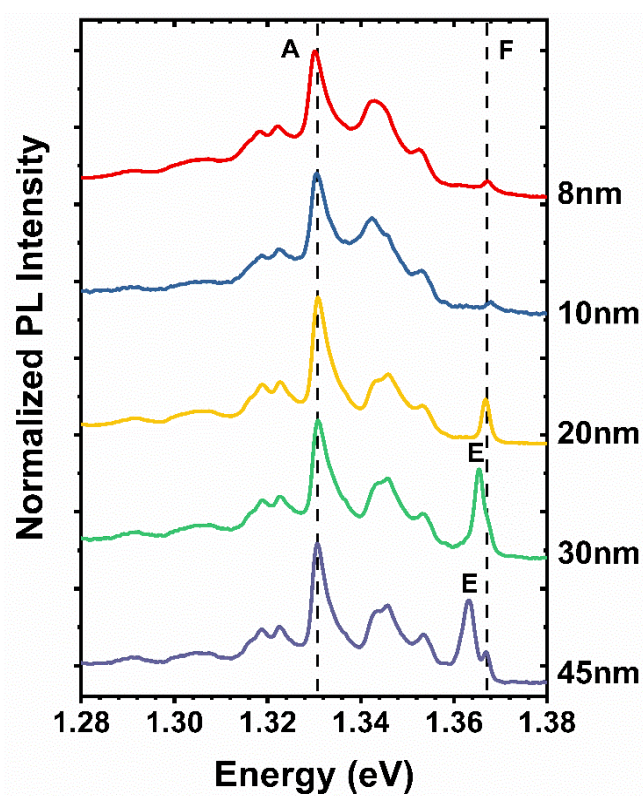

**Figure S1.** PL spectrum obtained at 4 K for 8nm, 10nm, 20nm, 30nm, and 45nm thick CrSBr

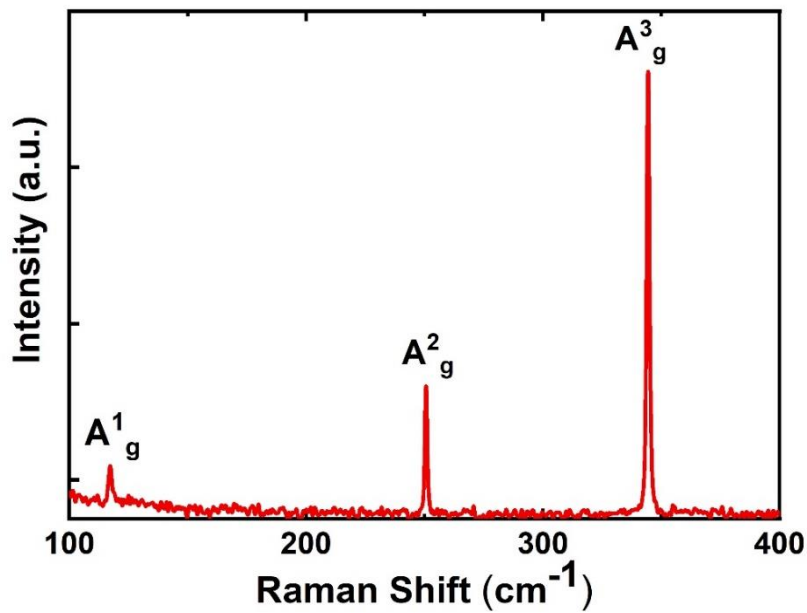

**Figure S2.** Raman spectrum for 532 nm excitation at 4 K. The Raman spectrum is measured with excitation laser polarized along the b axis, showing the out-of-plane $A^1_g$ mode ~ 118 cm$^{-1}$ (14.6 meV), $A^2_g$ mode ~ 251 cm$^{-1}$ (31.1 meV) and $A^3_g$ modes ~ 345 cm$^{-1}$ (42.8 meV).

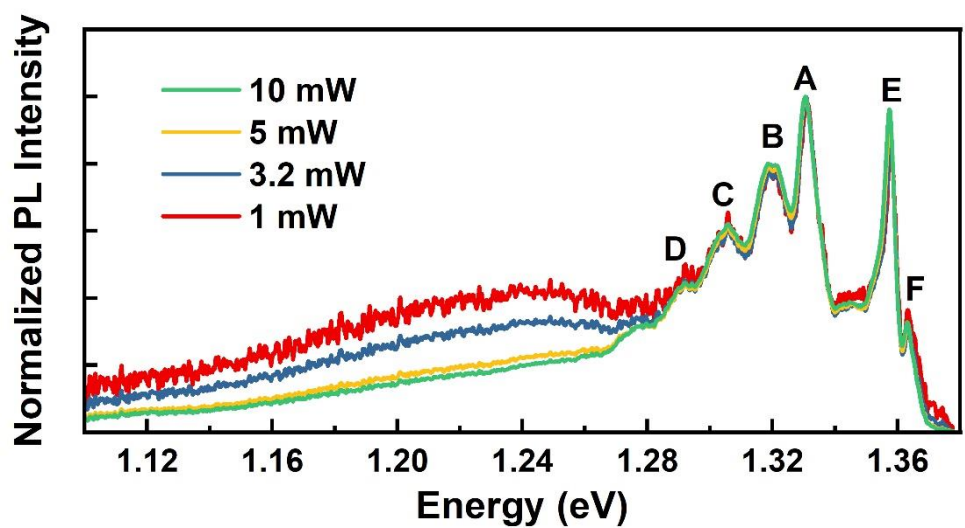

**Figure S3.** Normalized Power-dependent PL spectra for 532 nm excitation at 4 K. The multi-peaks remain proportionally unchanged with increasing pumping power, indicating that the broad PL spectrum contains and only contains exciton-phonon and exciton-polariton states.

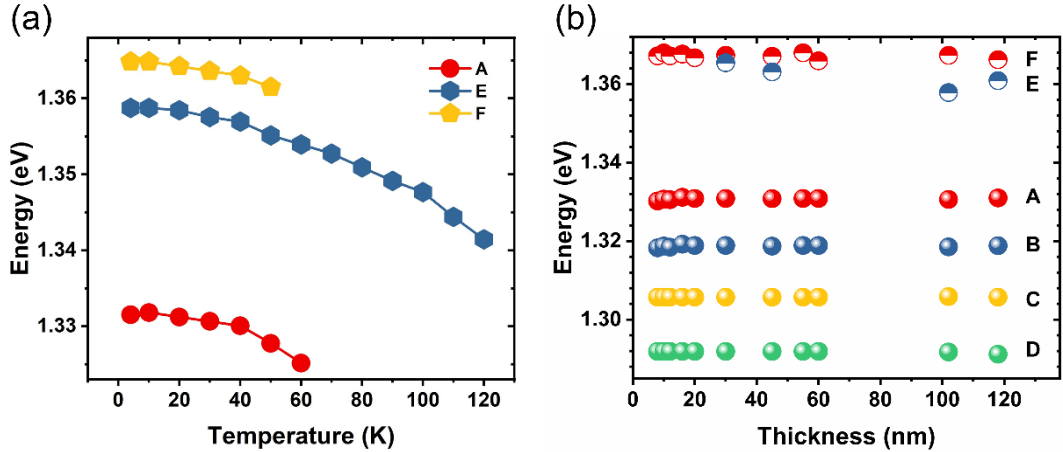

**Figure S4.** (a) Evolution of peak A, E and F from 4 K to 120 K. (b) Energy of peaks in PL spectra obtained at 4 K for CrSBr as a function of thickness. The circles are energies obtained from Lorentzian fitting.

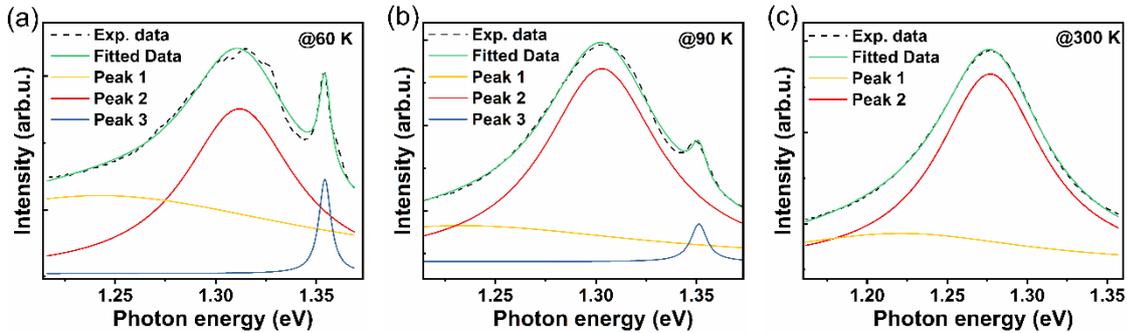

**Figure S5.** PL spectrum and Lorentzian Fitting curves of CrSBr acquired at (a) 60 K (b) 90 K and (c) 300 K.

The data presented in Figure 3 of the main text were obtained by fitting the experimental PL spectra with Lorentzian functions. It shows the FWHM and peak position of the main emission across the temperature range of 60 K to 300 K, and integrated PL intensity from 4 K to 300 K. Figure S5 demonstrate this fitting procedure for three representative PL spectra acquired at 60 K, 90 K and 300 K. Notably, the PL data below 60 K were not included in the analysis of peak position and FWHM due to the distinct splitting in different phonon sideband emissions. Nevertheless, the collected data from 60 K to 300 K are adequate for elucidating the exciton-phonon coupling mechanism in the investigated AFM CrSBr, with a Néel temperature ($T_N$) at around 135 K. For PL spectra taken at 60 K and 70 K, we approximated the data with a broad peak encompassing individual phonon-sideband emissions. Spectra with a distinct peak at approximately 1.35 eV were fitted with three Lorentzian functions: the first for background compensation, the second representing the main emission from exciton-phonon coupling, and the third (identified as E in the main text) linked to exciton-polariton states from strong coupling between excitons and photons in our 102 nm-thick CrSBr crystal. PL spectra above the $T_N$ (above 140K) were fitted using two Lorentzian functions. As for the integrated PL intensity across the temperature range of 4 K to 300 K, it was the integration of the fitted Lorentzian functions representing the exciton-phonon coupling branch at each temperature.

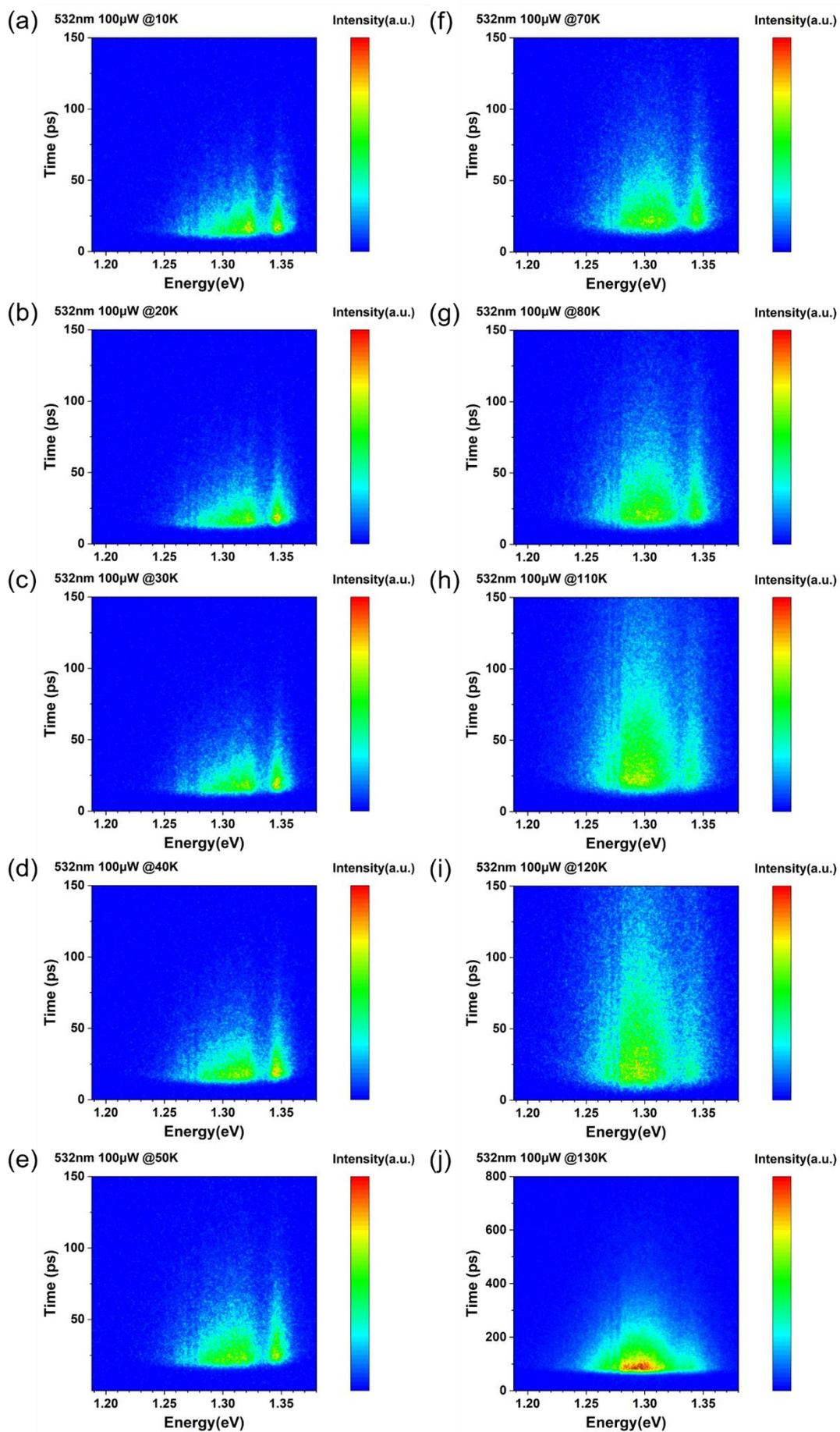

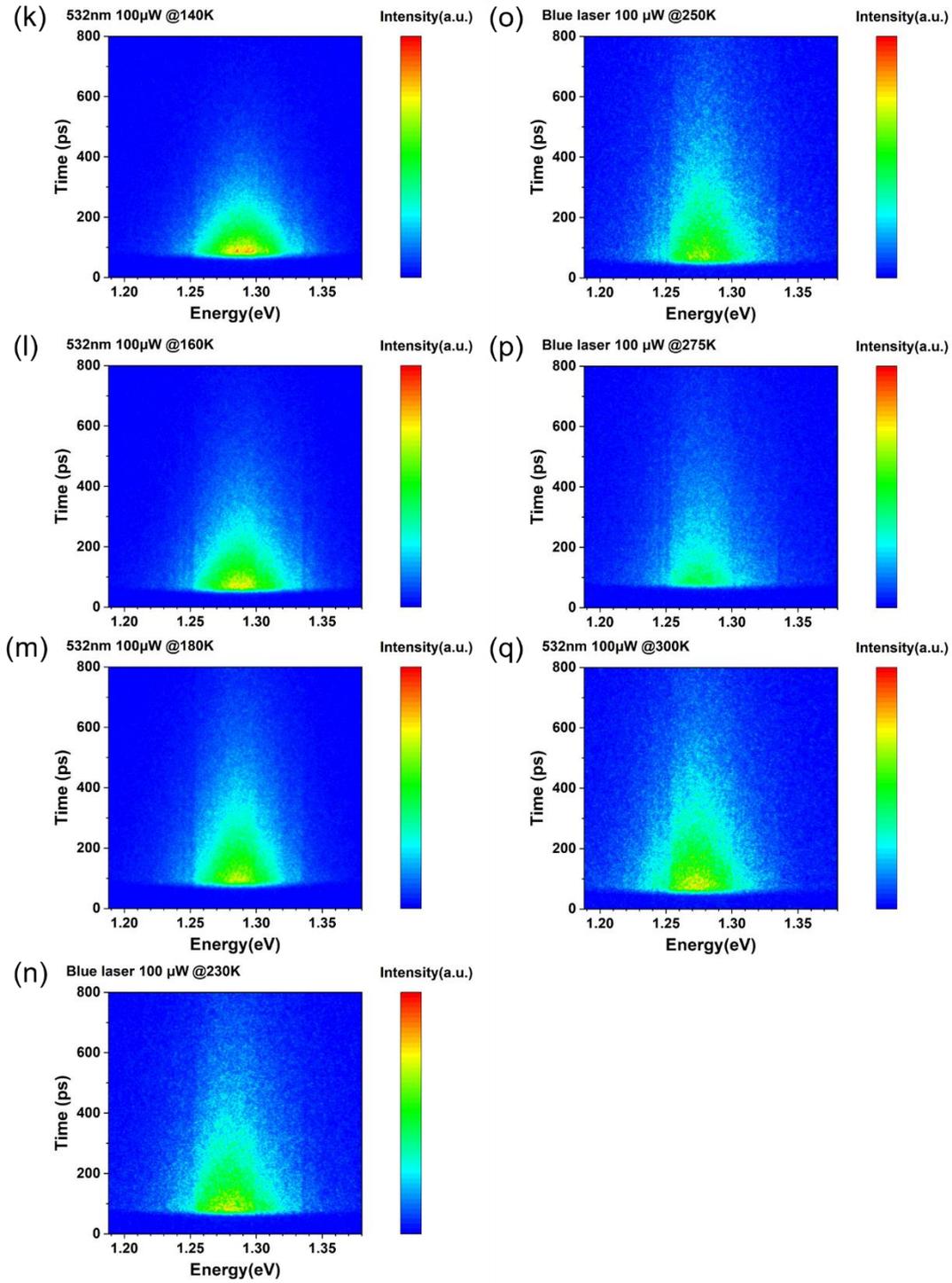

**Figure S6.** Spectrally-resolved streak-camera image of the PL at temperatures of **(a)** 10 K, **(b)** 20 K, **(c)** 30 K, **(d)** 40 K, **(e)** 50 K, **(f)** 70 K, **(g)** 80 K, **(h)** 110 K, **(i)** 120 K, **(j)** 130 K, **(k)** 140 K, **(l)** 160 K, **(m)** 180 K, **(n)** 230 K, **(o)** 250 K, **(p)** 275 K, **(q)** 300 K.